\documentclass[
  ]
  {aipproc}

\layoutstyle{6x9}
\usepackage{graphicx}
\DeclareGraphicsExtensions{.eps}
\usepackage{subfigure}

\newcommand{\mathvec}[1]{\mbox{\boldmath$#1$}}

\begin{document}

\title{Probing the electron EDM with cold molecules}


\author{B.~E. Sauer, H.~T. Ashworth, J.~J. Hudson, M.~R. Tarbutt, E.~A. Hinds}{
address={Centre for Cold Matter,
Imperial College London, London, SW7 2BW, UK\\
e-mail: ed.hinds@imperial.ac.uk}
}

\begin{abstract}
We present progress towards a new measurement of the electron
electric dipole moment using a cold supersonic beam of YbF
molecules.  Data are currently being taken with a sensitivity of
$10^{-27}\textrm{e.cm}/\sqrt{\textrm{day}}$. We therefore expect to
make an improvement over the Tl experiment of Commins' group, which
currently gives the most precise result.  We discuss the systematic
and statistical errors and comment on the future prospect of making
a measurement at the level of
$10^{-29}\textrm{e.cm}/\sqrt{\textrm{day}}$.
\end{abstract}

\maketitle


\section{Motivation and historical perspective}

The most precise electric dipole moment (EDM)
measurement~\cite{Commins02} on the electron gives $d_e = (6.9 \pm
7.4)\times 10^{-28}\ \textrm{e.cm}$, a result consistent with zero.
A non-zero result at this level would necessarily mean the violation
of time reversal symmetry (T) symmetry and the existence of new
particle physics, beyond the Standard Model~\cite{Fortson03}.
Although the Standard Model does exhibit T violation through the CKM
mechanism, the electron EDM produced by this mechanism is roughly
ten orders of magnitude below the present experimental sensitivity
and is for all practical purposes zero. By contrast, many modern
extensions of particle theory lead quite naturally to a value in the
current range of $10^{-27}\ \textrm{e.cm}$ or a little
below~\cite{CPVWS}. Since our experiment using cold YbF molecules
aims to be more sensitive than this, it is a search for new physics.
Assuming the validity of CPT, an electron EDM at the $10^{-27}\
\textrm{e.cm}$ level would also imply a new type of CP violation.
This would be of great interest for cosmology as it seems to be a
necessary ingredient in understanding the observed matter-antimatter
asymmetry of the universe~\cite{Sakharov}.

\section{Principles of the experiment}

The interaction between $d_e$ and an applied field $\bf{E}$ can be
expressed by the effective non-relativistic Hamiltonian  $ -
d_e\,\alpha(E)\,\hat{\mathvec{\sigma}} \cdot \bf{E}$. For a free
electron, $\alpha(E) = 1$ and $\hat{\mathvec{\sigma}}$ is a unit
vector along the spin. If the electron is part of an atom or
molecule, $\hat{\mathvec{\sigma}}$ lies along the spin of the system
and $\alpha(E)$ is a factor that depends on the structure. Some
heavy atoms and molecules have the virtue that $\alpha(E)\gg1$, and
then it is called the enhancement factor~\cite{Sandars:atom}. This
coupling resembles the interaction
$-\mu\,\beta(B)\,\mathvec{\hat{\sigma}} \cdot \bf{B}$ of the
magnetic moment $\mu$ with a magnetic field $\bf{B}$, where
$\beta(B)$ accounts for the atomic or molecular structure. It is
instructive to compare these two interactions in the case of a free
electron with an EDM of, say, $d_e = 5\times10^{-28}\
\textrm{e.cm}$, just below the present limit. In a
$100\,\textrm{kV/cm}$ field the EDM energy is so small that it
equals the magnetic energy in a field of only $9\times10^{-19}\
\textrm{T}$. Controlling the stray magnetic field at that level
seems close to impossible, especially when applying the electric
field. Heavy atoms such as Cs and Tl alleviate this problem by their
large enhancement factors. In particular, $\alpha(E) = -585$ for the
thallium atom~\cite{Liu}, which relaxes the necessary field control
to the challenging, but achievable fT level. Two magnetic effects
are most troublesome. (i) Stray magnetic fields vary both in space
and time. (ii) Atoms moving through the large electric field
experience a motional magnetic field~\cite{AC_paper} $\bf{E\times
v}/c^2$. In both cases the unwanted field components are typically
many orders of magnitude larger than 1\,fT and heroic efforts were
needed to reach the current precision~\cite{Commins02}.

\begin{figure}[t]
\centering
\includegraphics[height=4cm]{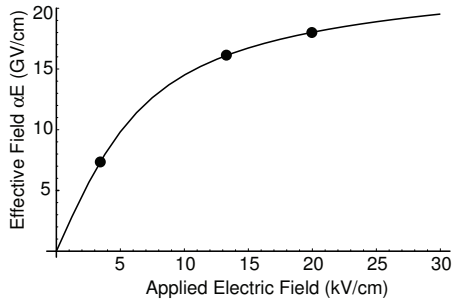}
\caption{Enhancement factor for YbF. The dots indicate field values
at which we have recorded EDM data.} \label{polarisation}
\end{figure}

Heavy polar molecules offer substantial relief from these
difficulties~\cite{physica_scripta}. First, the enhancement factors
are generically much larger~\cite{molecule_enhancement} because the
electron EDM interacts with the polarisation of the charge cloud
close to the heavy nucleus. In an atom this polarisation follows
from the mixing of higher electronic states by the applied electric
field. In a polar molecule, these electronic states are already
strongly mixed by the chemical bond and it is only rotational states
that have to be mixed by the applied field. Since these are
typically a thousand times closer in energy, the molecular
enhancement factor is correspondingly larger. For the YbF molecule
used in our experiment, the enhancement factor~\cite{Kozlov:1998} is
$\alpha \simeq 10^6$ at our operating field of $13\ \textrm{kV/cm}$,
which relaxes the requirement on field control to the $\textrm{pT}$
level. Figure \ref{polarisation} shows the enhancement (expressed as
an effective electric field) as a function of the laboratory field.
Note that the polarisation, due to rotational state mixing,
saturates at relatively modest fields.

There is a second advantage to YbF\footnote{or any system whose
tensor Stark splitting greatly exceeds the Zeeman interaction.}.
Being polar, this molecule has a strong tensor Stark splitting
between sublevels of different $|m|$, the total angular momentum
component along the electric field direction. This strongly
suppresses the Zeeman shift due to perpendicular magnetic fields,
including the motional field $\bf{E\times v}/c^2$. For our typical
operating parameters, the motion-induced false EDM is reduced by
this mechanism to a level below
$10^{-33}\,\textrm{e.cm}$~\cite{Hudson}, which is negligible.

\begin{ltxfigure}[htb]
\centering
\subfigure[]{\includegraphics[width=6.5cm]{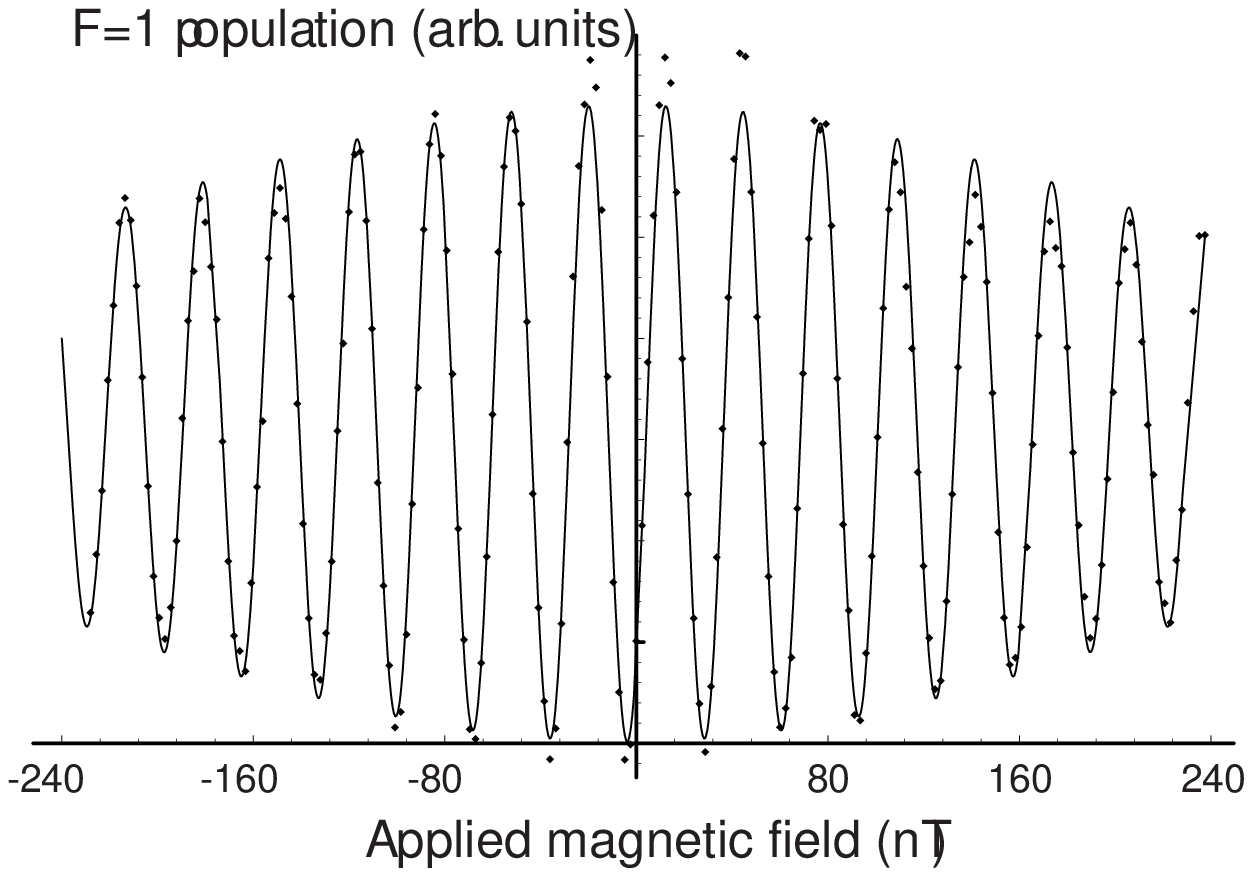}\label{fringes}}\quad
\subfigure[]{\includegraphics[width=6cm]{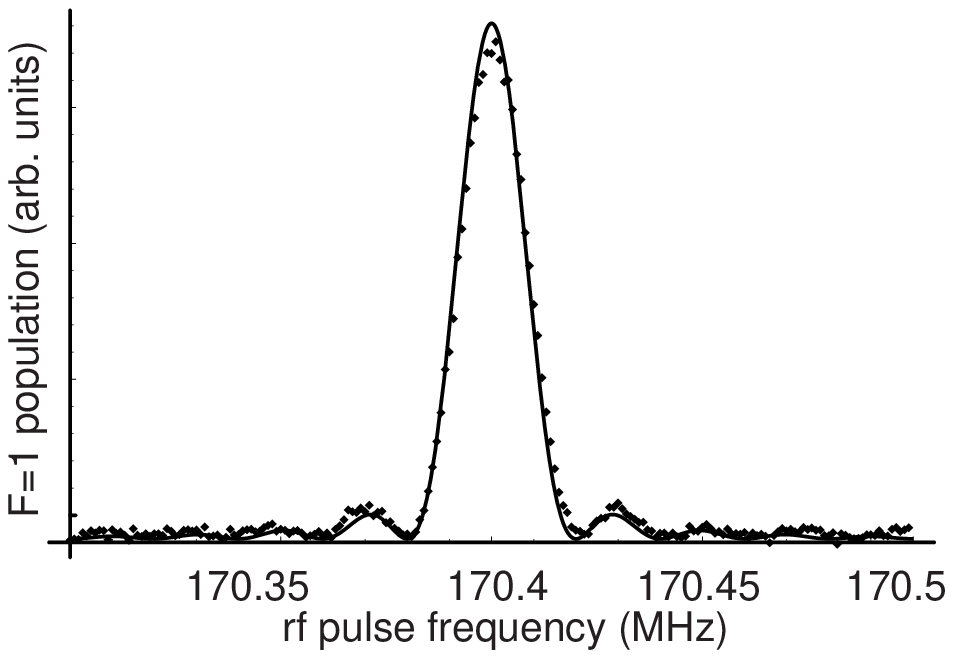}\label{rf_lineshape}}
\caption{a) YbF interferometer fringes. Dots: $F=1$ population
measured by fluorescence. Curve: Calculated fringes with
normalisation and magnetic field offset as free parameters.
Variation of fringe visibility is due to the known beam velocity
distribution. b) Lineshape for rf beam splitter. Dots: experiment.
Line: model with 50$\mu$s rf pulse length.}
\end{ltxfigure}

Our experiment uses a cold, pulsed, supersonic beam of YbF
radicals~\cite{Tarbutt} in a magnetically shielded vertical vacuum
chamber $\sim 1.5$m high. The electronic, rotational and vibrational
ground state $X^2\Sigma^+_{1/2},N=0,v=0$ is a hyperfine doublet with
states $F=0,1$ split by 170\,MHz. We first deplete the $F=1$ state
by laser excitation of the $F=1$ molecules on the $A_{1/2}\leftarrow
X$ transition. This laser beam is called the pump. A radiofrequency
magnetic field, which we call the first beam splitter, then drives
the $F=0$ molecules into a symmetric coherent superposition of the
$F=1, m_F = \pm 1$ states, as described later in more detail. Next,
parallel dc electric and magnetic fields are applied to introduce a
phase shift $\Delta\phi = \frac{2}{\hbar}\int_0^\tau ( d_e\
\alpha(E)\ E(t) + \mu\ \beta(B)\ B(t)) \mathrm{d}t $ between the two
superposed states. Here $E$ and $B$ appear as functions of time
because they are the fields in the molecular rest frame. At time
$\tau$ the molecules interact with a second oscillating field, the
recombining beam-splitter, that couples the symmetric part of the
$F=1$ coherence back to the $F=0$ state. The resulting $F=0$ state
population exhibits the usual $\cos^2\left(\Delta\phi/2\right)$
fringes of an interferometer. We detect the complementary $F=1$
population using fluorescence induced by a probe laser on the
$A_{1/2}\leftarrow X$ transition. Figure\,\ref{fringes} shows the
interference fringes observed in this fluorescence when the magnetic
field is scanned.

The beam splitter is an rf magnetic field perpendicular to $\bf{E}$
and along the beam direction.  When a pulse of molecules arrives at
the centre of the rf loop it is subject to a short pulse of resonant
$170\,$MHz radiation, which induces hyperfine population transfer.
It is important that the molecules move as little as possible during
this transition because unwanted phase shifts can occur if the
fields $E(t)$ or $B(t)$ rotate during the splitting (or recombining)
transition. When combined with other imperfections of the apparatus,
such a phase can produce a false EDM and is therefore undesirable.
Figure\,\ref{rf_lineshape} shows the lineshape measured using a
$50\,\mu$s-long rf pulse, together with a fit to the standard Rabi
formula\cite{icols} showing good agreement. With high power
amplifiers, we are now able work with pulses $10\,\mu$s long,
corresponding to a beam movement below 6\,mm. The peak transition
probability in Figure\,\ref{rf_lineshape}, taken using the first
loop, is approximately 0.8. This is due to the beam velocity spread,
which gives the gas pulse a length of 5\,cm at the first loop and
10\,cm at the second. Since the rf field strength varies along the
beam line, this spatial spread produces a small distribution of Rabi
frequencies, which could be improved by having a more monoenergetic
beam.

The arrival of each YbF pulse at our detector is recorded with
$1\mu\textrm{s}$ resolution. This time-resolved data gives us a
spatial resolution of $\sim 5\ \textrm{mm}$ at the second rf loop.
This is the basis of a useful diagnostic technique: varying the
timing of the rf pulse allows us to map out the field distribution
in the apparatus. This works very well for electric fields because
the Stark shift of the hyperfine transition is large. It can also be
used to probe magnetic fields, albeit with slightly lower spatial
resolution.

\section{Noise and systematic effects}

With the YbF interferometer working as described above, the electron
EDM measurement is straightforward in principle. A small magnetic
field is applied to bias the interferometer phase to the point of
highest slope. The applied electric field is then reversed and the
change in the interferometer output constitutes the EDM signal. In
practice, we also reverse the direction of the magnetic field and
modulate its amplitude. The EDM is correlated with the relative
direction of E and B while the signals demodulated on the individual
switching channels yield valuable information about the drift of the
interferometer setup and some systematic effects. Additional
modulations can be applied: the relative phase between the splitter
and recombiner rf fields, the laser frequency, the rf amplitudes and
the rf frequencies.

\begin{ltxfigure}[b]
\centering
\subfigure[]{\includegraphics[width=6cm]{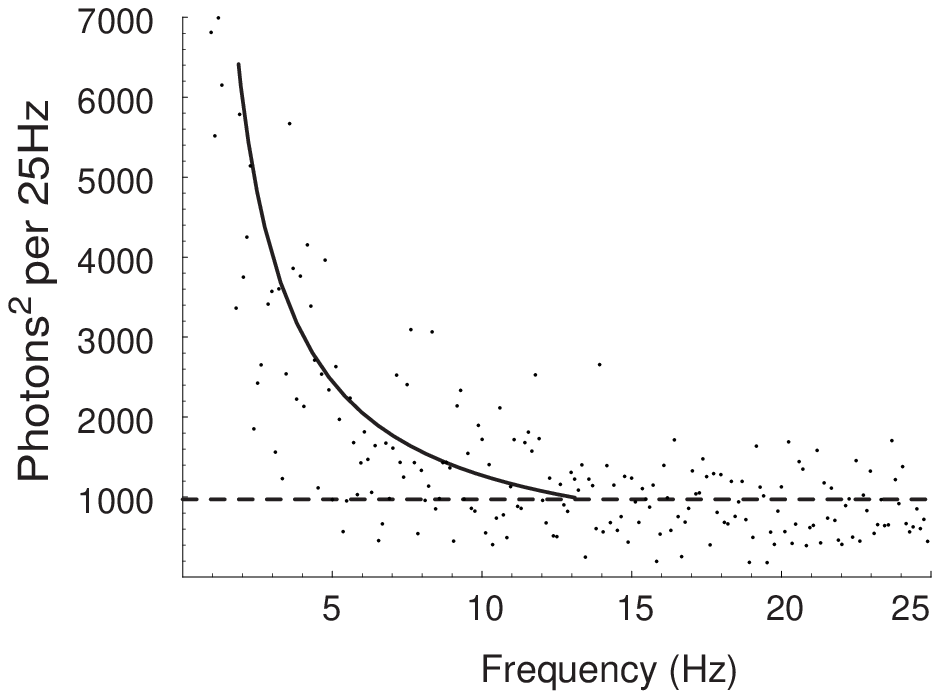}\label{noise_a}}\quad
\subfigure[]{\includegraphics[width=6cm]{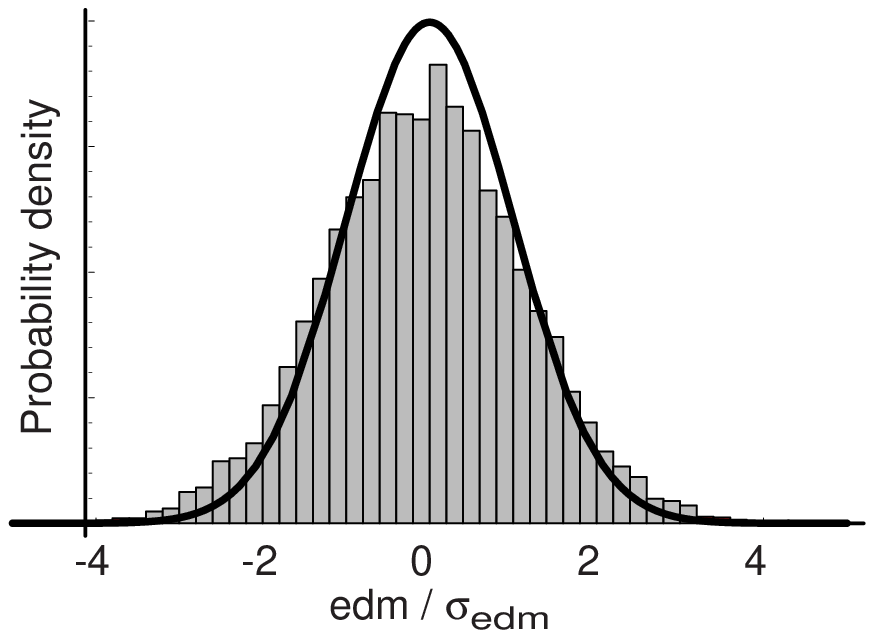}\label{noise_b}}
\caption{a) Measured noise power spectrum. Solid line: $1/f$ noise,
 Dashed line: expected shot noise level. b) Distribution of
 some 9500 individual EDM measurements, each normalised to its own standard deviation.
 The solid curve is a Gaussian of unit width. There are clearly excess points in the wings,
 producing a corresponding deficit at the centre. These are due to the non-statistical behaviour
 of the noise.} \label{noise_plots}
\end{ltxfigure}

With four switching channels (e.g. E reversal, B reversal, relative
rf phase and a small step $\Delta B$) there are sixteen possible
states of the machine. We cycle through all of them 256 times to
form a ``block'' of data consisting of 4096 individual points. The
repetition rate is determined by the 25Hz ablation laser which forms
the YbF beam, thus each block of data requires about $3\,$minutes of
real time to record. Slow drifts in the signal are cancelled by
choosing appropriate modulation waveforms for the switching
channels~\cite{harrison}. Figure \ref{noise_a} shows the power
spectrum of the noise in our molecular beam signal, as measured by
the photomultiplier recording laser-induced fluorescence from the
$F=1$ state.  Above about 10Hz the spectrum is largely flat, as one
would expect for shot noise due to the statistics of the photon
arrivals, but at lower frequencies $1/f$ noise dominates. For this
reason, the noise we observe in our EDM measurement usually exceeds
the shot noise level. When the ablation target is moved to expose a
fresh spot to the YAG laser the excess is typically a factor of 1.5,
increasing to 2 over the course of an hour or so, at which point we
reset the target.

This non-statistical contribution to the noise in the EDM
measurement generates excess noise in the wings of the normalized
distribution shown in figure \ref{noise_b}. Although it is a small
effect, this must be taken into account if we are to assign reliable
confidence limits to our EDM measurement. We use a novel bootstrap
method. The basic idea is that the experiment itself provides the
best available estimate of the true underlying probability
distribution~\cite{bootstrap}. Each block of data provides a
measurement of the EDM.  By randomly choosing values of the EDM from
this experimental data set, we create an ensemble of additional
synthetic EDM data sets having the same distribution. Note that
chosen points are not removed from the pool. Statistical averages
are then performed using these empirical data sets. This method
works without needing any analytic model for the probability
distribution; in particular, we avoid making the usual assumption of
a normal distribution. Figure \ref{bootstrap} shows the bootstrapped
sampling distribution for our 13kV/cm data set and the corresponding
cumulative distribution. One sees that the bootstrap method is
putting excess probability into the wings of the distribution, as
expected. The integral of this probability distribution tells us
that the true confidence limits are about 20\% larger than one would
derive from applying normal Gaussian statistics to the data set.

\begin{ltxfigure}[t]
\centering
\subfigure[]{\includegraphics[width=5cm]{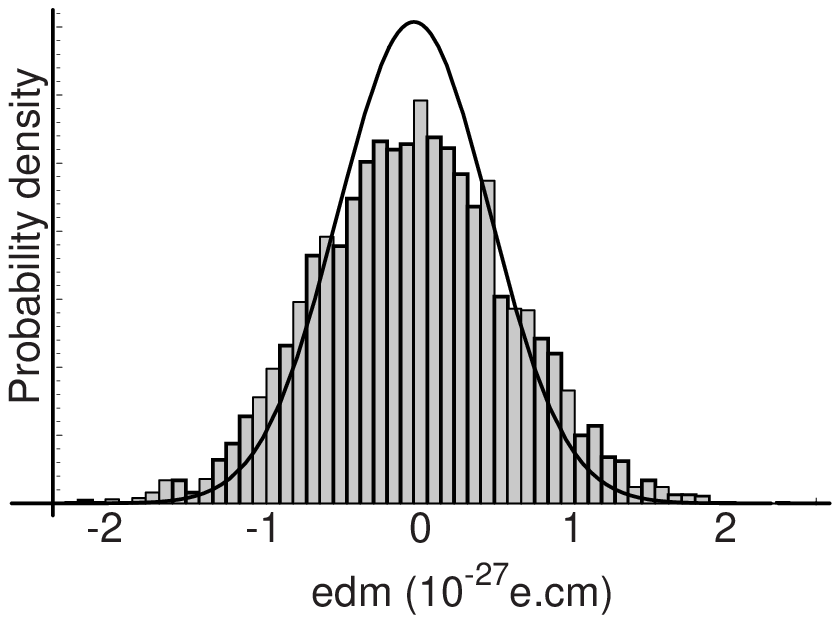}\label{boot_a}}\quad
\subfigure[]{\includegraphics[width=5cm]{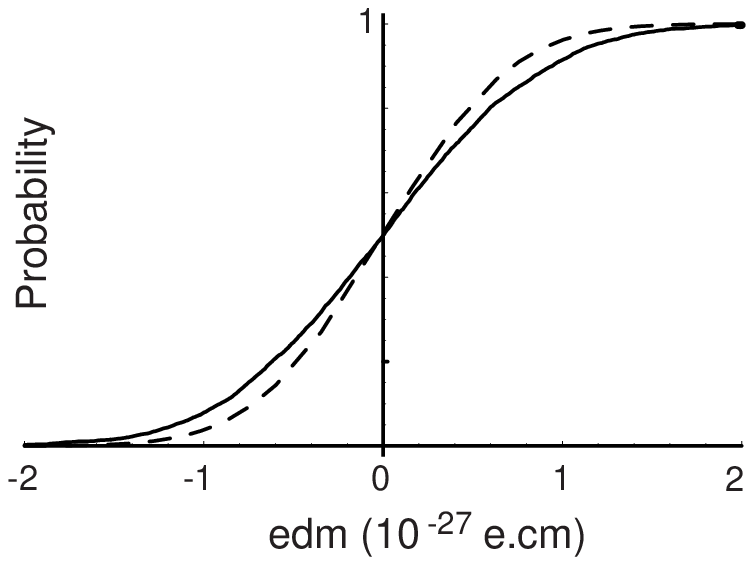}\label{boot_b}}
\caption{The bootstrap analysis. a) Bootstrapped EDM probability
distribution around the mean. The solid curve is a normal
distribution. b) Cumulative probability, integrating up from below
the mean for normal (dashed) and bootstrapped (solid) statistics.}
\label{bootstrap}
\end{ltxfigure}

The fluctuating magnetic field in the laboratory is a possible
source of additional noise in the measured EDM (or worse, a
systematic error if changes in the laboratory field are correlated
with the switching of the electric field). The YbF beam is protected
from external fields by two layers of magnetic shielding; a flux
gate magnetometer between the shields monitors the residual field
during data collection. Figure \ref{bnoise_a} shows the correlation
between the field measured by the molecular beam and by the
magnetometer, both multiplied by $\mu_B/(\alpha E)$ to express them
in units of EDM. The slope of the line shows that the shielding
factor of the inner shield is approximately $100$. As there is no
correlation between the external B field and the electric field
reversal, we are confident that this magnetic field noise does not
generate a false EDM at the present level of sensitivity.

We employ a veto which discards data when the external field noise
exceeds a certain threshold. This improves the uncertainty in the
EDM measurement, as illustrated in figure \ref{bnoise_b} for a
subset of our data. This plots the 67\% confidence limit on the EDM
vs.
 the magnetic field veto level. The optimal cutoff level is approximately where the external
  magnetic field induces fluctuations comparable to the EDM measurement error bar. A veto
  at this level typically rejects about 15\% of our data.

\begin{ltxfigure}[t]
\centering
\subfigure[]{\includegraphics[width=6cm]{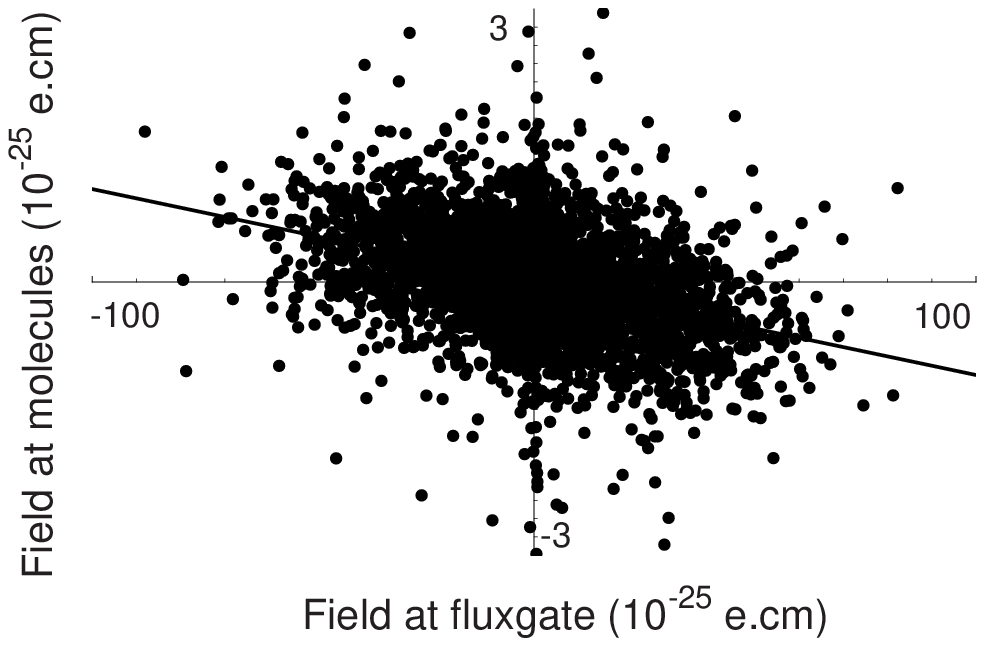}\label{bnoise_a}}\quad
\subfigure[]{\includegraphics[width=6cm]{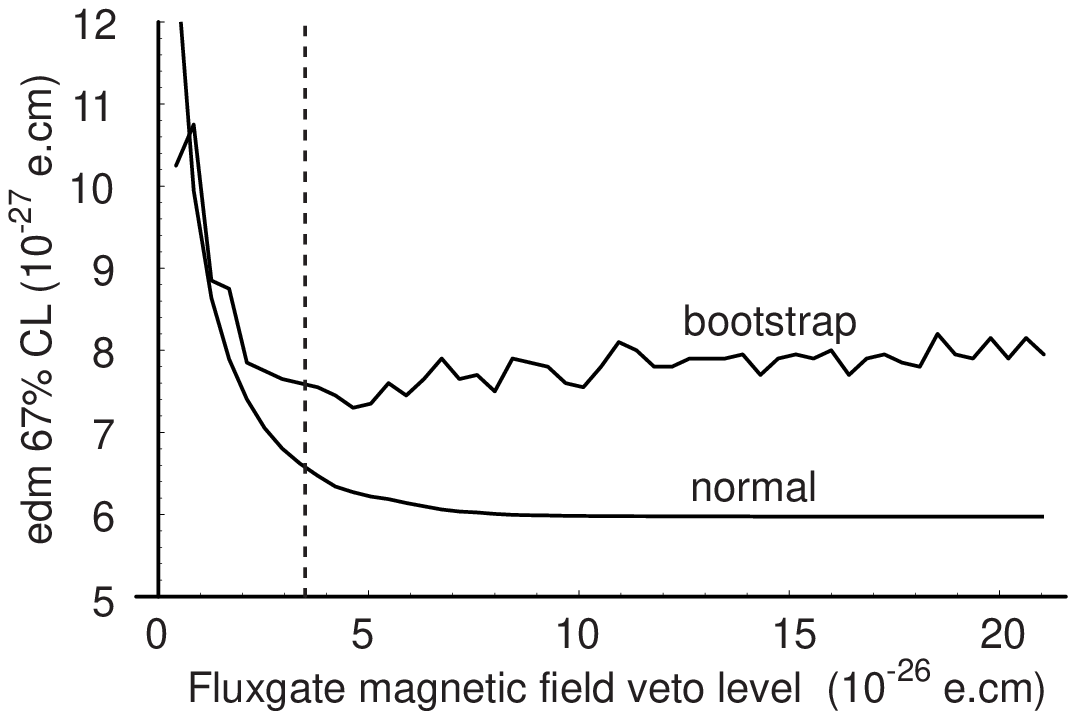}\label{bnoise_b}}
\caption{Magnetic noise. a) Correlation between the magnetic field
measured by molecules vs. field monitored externally. b) EDM
confidence limits as a function of magnetic veto for normal and
bootstrapped statistics. The dashed line shows the analysis veto
level.} \label{b_noise}
\end{ltxfigure}

During EDM data acquisition runs, we periodically reverse the
electric and magnetic connections to the apparatus manually. This
guards against a false EDM generated by the external apparatus (for
example magnetic fields from the high voltage relays) or the data
acquisition electronics. Because the experiment has been carefully
designed to minimize such effects, we do not find any signal
correlated with these manual reversals. The saturation of the
enhancement shown in figure \ref{polarisation} is another powerful
discriminant against systematics.  The interferometer phase induced
by a true EDM would have to vary in this way, whereas a systematic
error would be unlikely to do so. For example, the phase shift due
to the magnetic field of a relay is constant, whilst the effect of
leakage currents or electrical breakdown grows linearly or faster
with the applied voltage. We have recorded data at the field values
indicated in the figure. From the data taken so far, we can already
see that we do not have any strong disagreement with the result of
reference~\cite{Commins02}, however we have discovered a new
systematic effect that needs to be addressed before we can
confidently give a new result at a higher level of accuracy.

\begin{figure}[t]
\centering
\includegraphics[height=4cm]{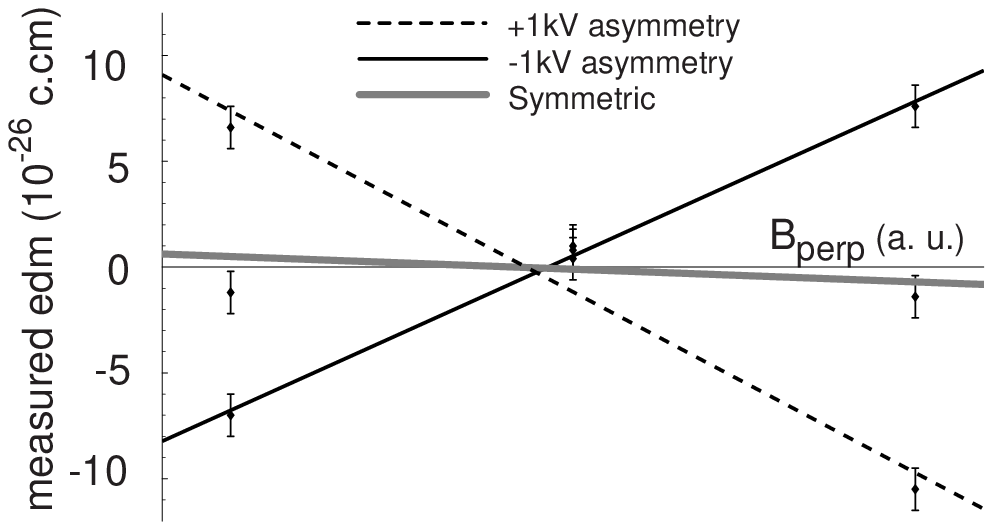}
\caption{False EDM due to asymmetric field rotation in a
perpendicular magnetic field.} \label{systematic}
\end{figure}

As noted previously, the tensor Stark splitting causes the Zeeman
shift to depend only on the magnetic field component $B_z$ along the
electric field direction. As a result, false EDMs due to the
$\textrm{v} \times E$ motional field and to geometric phases induced
by rotating magnetic fields are utterly negligible for the current
beam experiment. This strong coupling of the molecular axis to the
external electric field is a major blessing, however, it does also
bring with it a potential problem. If the direction of the electric
field does not reverse exactly when the voltages are switched, this
change of z-axis can produce a change of $B_z$ and hence a Zeeman
shift that mimics an EDM. In our apparatus, the electric field
plates are split into three regions so that the splitter and
recombiner transitions can take place at a lower field than the
central field region of the interferometer. In the gaps between
these regions, the field lines are curved, causing off-axis
molecules to experience a rotation of the local z-axis as they fly
through. If the electric field were to reverse perfectly this would
not cause a problem since the rotation angle would be the same for
both electric polarities, but of course no reversal can be perfect.
Consequently, a fixed magnetic field roughly perpendicular to the
rotating electric field can induce a false EDM. We have demonstrated
this by applying a strong perpendicular magnetic field and
deliberately introducing a very large asymmetry when the electric
field reverses. Figure \ref{systematic} shows a large false EDM
induced in this way. Under our normal operating conditions, we
estimate that this effect should not be larger than $\sim 10^{-28}\
\textrm{e.cm}$.

Because it is difficult to guarantee that we understand these fringe
fields perfectly and because we are aiming to measure $d_e$ at the
$\sim 10^{-28}\ \textrm{e.cm}$ level of accuracy, we are now
replacing the electrode structure with a single pair of field plates
so that the molecules will experience the same magnitude and
direction of electric field throughout the entire interferometer. In
the new high-field splitter and recombiner regions, the field must
be homogeneous across the beam to within 0.1\% in order to avoid
excessive Stark broadening of the rf transition. This places severe
limits on the parallelism of the plates but the required precision
has been achieved with specialist machining techniques and the new
plates are being installed at time of writing.

\section{Outlook}
At present we anticipate making a measurement with accuracy in the
$10^{-28}\textrm{e.cm}$ range as soon as the new plates are working,
which will improve on the result of ref.~\cite{Commins02}. Over the
next few years we expect to reach an accuracy of $1 \times
10^{-28}\textrm{e.cm}$ without additional major modifications. In
order to be sure of systematics at this level, it may also be
necessary to perform a control experiment using CaF molecules. These
are similar to YbF structurally and magnetically, but have $\sim 40$
times less sensitivity to the electron EDM according to the expected
$Z^3$ scaling~\cite{physica_scripta}. We have already made a cold
supersonic CaF beam in our apparatus and have seen the transitions
that are needed.

It seems possible to go significantly further in measurement
accuracy by decelerating the YbF molecules to increase the time that
they spend in the interferometer.  With this in mind, we have
demonstrated that our YbF beam can be decelerated~\cite{decelerator}
and we have made substantial progress in understanding how to bring
heavy polar molecules close to rest using an alternating gradient
decelerator~\cite{AG review}. With the use of an intense, slow YbF
beam, there is no obvious obstacle to a measurement at the level of
$1 \times 10^{-29}\textrm{e.cm}$. Such high precision would provide
a probe of new CP-violating elementary particle physics up to mass
scales in the range of many TeV, testing some models far beyond the
reach of current accelerators including the LHC at CERN.


\begin{theacknowledgments}
We acknowledge support from PPARC, EPSRC, the Royal Society (UK), and
the Cold Molecules Research Training Network of the European
Commission. We are indebted to Jon Dyne for expert technical
assistance.
\end{theacknowledgments}






\begin{thebibliography}{99}

\bibitem{Commins02} B.C. Regan \emph{et al.}, \emph{Phys. Rev. Lett.} \textbf{88}, 071805
(2002).

\bibitem{Fortson03} E. N. Fortson, P. Sandars, and S. Barr, Phys. Today \textbf{56}, No. 6, 33 (2003).

\bibitem{CPVWS} I.B. Khriplovich and S.K. Lamoreaux, \emph{CP
violation without strangeness}. (Springer, Berlin 1997); Maxim
Pospelov and Adam Ritz, Annals Phys. \textbf{318}, 119-169 (2005).

\bibitem {Sakharov} A.D. Sakharov, \emph{Pis'ma ZhETF} \textbf{5}, 32 (1967).
[\emph{Sov. Phys. JETP Lett.} \textbf{5}, 24 (1967)]; M Dine, A
Kusenko, Rev. Mod. Phys. \textbf{76}, 1 (2004).

\bibitem{Sandars:atom} P.G.H. Sandars, \emph{Phys. Lett.} \textbf{14}, 194 (1965).

\bibitem{Liu} Z.W. Liu and H. P. Kelly, \emph{Phys. Rev. A} \textbf{45}, R4210 (1992).

\bibitem{AC_paper} Karin Sangster \emph{et al.}, \emph{Phys. Rev. Lett.} \textbf{71}, 3641
(1993); \emph{Phys. Rev. A.} \textbf{51}, 1776 (1995).

\bibitem{physica_scripta} E. A. Hinds, \emph{Physica Scripta} \textbf{T70}, 34
(1997).

\bibitem{molecule_enhancement}P.G.H. Sandars in \emph{Atomic Physics
4} ed. G. zu Putlitz, (Plenum, 1975) p.71.

\bibitem{Kozlov:1998} M.G. Kozlov and V.F. Ezhov, \emph{Phys. Rev.} \textbf{A49}, 4502 (1994);
M.G. Kozlov, \emph{J. Phys. B} \textbf{30} L607 (1997); A. V.
Titov, N. S. Mosyagin, V. F. Ezhov, \emph{Phys. Rev. Lett.}
\textbf{77} 5346 (1996); H.M. Quiney, H. Skaane, I.P. Grant,
\emph{J. Phys. B} \textbf{31} L85 (1998)(after correcting for the
trivial factor of 2 between \textbf{s} and \textbf{$\sigma $}
their result becomes 26~GV/cm); F.A. Parpia, \emph{J. Phys. B}
\textbf{31} 1409 (1998); N. Mosyagin, M. Kozlov, A. Titov,
\emph{J. Phys. B} \textbf{31} L763 (1998).

\bibitem{Hudson} J. J. Hudson \emph{et al.}, \emph{Phys. Rev.
Lett.} \textbf{89}, 023003 (2002).

\bibitem{Tarbutt} M. R. Tarbutt \emph{et al.}, \emph{J. Phys. B} \textbf{35}, 5013 (2002).

\bibitem{icols}J. J. Hudson, P. C. Condylis, H. T. Ashworth, M. R. Tarbutt, B. E. Sauer and E. A Hinds, in \emph{Proceedings of the 17th Conference on Laser Spectroscopy} (World Scientific, Singapore 2005), p.129-136

\bibitem{Ramsey} Norman F. Ramsey, \textit{Molecular beams}, (Oxford
University Press, 1956).

\bibitem{stark_paper} P. C. Condylis \emph{et al.}, J. Chem. Phys. \textbf{123} 231101 (2005).

\bibitem{harrison} G.~E. Harrison, M.~A. Player, and P.~G.~H. Sandars, \emph{J. Phys.
E} \textbf{4}, 750 (1971).

\bibitem{bootstrap} B. Efron and R. Tibshirani, \emph{Stat. Sci.} \textbf{1}, 54 (1986).

\bibitem{decelerator} M.R. Tarbutt \emph{et al.}, \emph{Phys. Rev. Lett.}, \textbf{92}, 173002 (2004).

\bibitem{AG review} H. L. Bethlem \emph{et al.}, \emph{J. Phys. B.}, \textbf{39}, R263-R291 (2006).

\end{thebibliography}
\end{document}